\documentclass[twocolumn,footinbib]{revtex4-2}
\usepackage{latexsym,epsfig,amssymb, amsmath,nicefrac, xcolor}

\usepackage{hyperref}

\usepackage{amsfonts,amsthm} 
\usepackage{bm,bbm}
\usepackage{graphicx}
\usepackage{esint}
\usepackage{color}
\usepackage{braket}

\newcommand{\be}{\begin{equation}}
\newcommand{\ee}{\end{equation}}
\newcommand{\ba}{\begin{eqnarray}}
\newcommand{\ea}{\end{eqnarray}}
\def\bs{\begin{subequations}}
\def\es{\end{subequations}}
\def\a{\alpha}

\def\vp{\varphi}

\newcommand{\rf}[1]{(\ref{#1})}

\begin{document}

\title{{\Large ACT, SPT, and chaotic inflation}}

\author{Renata Kallosh${}^1$, Andrei Linde${}^1$ and Diederik Roest${}^2$}

\affiliation{{}$^1$Department of Physics and SITP, Stanford University,\\ Stanford, California 94305 USA}

\affiliation{{}$^2$Van Swinderen Institute for Particle Physics and Gravity, \\ University of Groningen, 9747 AG Groningen, The Netherlands} 

\begin{abstract}
We show that the simplest generalization of the chaotic inflation model $\tfrac12 {m^{2}\phi^{2}}$ with nonminimal coupling to gravity $(1+\phi) R$ provides a good match to the results of the latest data release of the Atacama Cosmology Telescope and South Pole Telescope, with $r \approx10^{{-2}}$.
\end{abstract}

\maketitle

The latest data release of the Atacama Cosmology Telescope (ACT)
 \cite{Louis:2025tst,ACT:2025tim} provides strong support for inflationary cosmology. However, its results, combined with those of the DESI DR1  data release \cite{DESI:2024uvr,DESI:2024mwx}, may significantly alter the constraints on the spectral index $n_{s}$. 
 
 The constraint given by Planck 2018 was  $n_{s } = 0.9651 \pm  0.0044$ \cite{Planck:2018vyg}. Over the years, there was a trend to higher values on $n_s$, see e.g.  \cite{Efstathiou:2019mdh} with the result $ n_s =  0.9683 \pm 0.0040$. Finally, a combination of Planck, ACT, and DESI-DR1  (P-ACT-LB) gives $n_{s} =0.9743 \pm  0.0034 $  \cite{Louis:2025tst}. This result differs from the original Planck result by about $2\sigma$.

\begin{figure}
\centerline{\includegraphics[scale=.14]{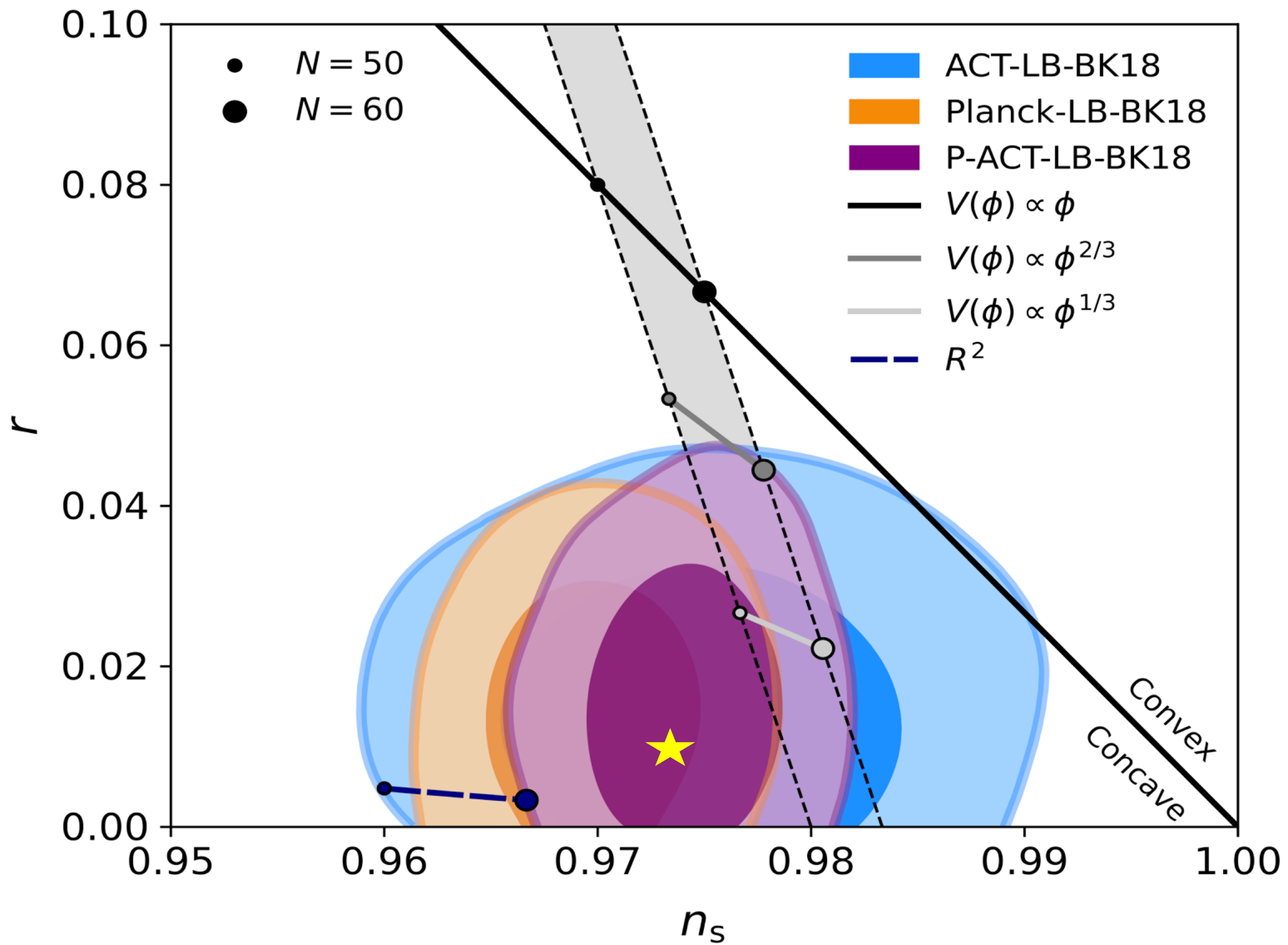}}
\caption{\footnotesize {The figure shows the latest constraints  on $n_{s}$ and $r$ according to ACT  (P-ACT-LB)  \cite{Louis:2025tst}. The dashed line at the bottom corresponds to the Starobinsky model. The yellow star at the center of the dark purple area favored by ACT shows the values of $n_{s}$ and $r$ \rf{60} in the  model  \rf{Jordan} for $N_{e}= 60$.}}
\vspace{-.3cm}
\label{xiStarAlone}
\end{figure}

The P-ACT-LB results are illustrated by Fig. 10 in \cite{ACT:2025tim}; we reproduce it here in our Fig. \ref{xiStarAlone}. The authors of \cite{ACT:2025tim} conclude, in particular, that the P-ACT-LB constraint of  $n_{s}$ disfavors the Starobinsky model \cite{Starobinsky:1980te} at $\gtrsim 2\sigma$.  A similar conclusion can be made with respect to the Higgs inflation \cite{Salopek:1988qh,Bezrukov:2007ep} and to many versions of $\alpha$-attractors \cite{Kallosh:2013yoa}.

This is a strong and rather unexpected statement. Therefore, in the original version of this paper, we noted that one should take it with caution, pending the release of the SPT data and the analysis of the recent DESI DR2 results.  

The new SPT-3G-D1 results \cite{SPT-3G:2025bzu} show that a combination of the Planck data and the latest ACT and SPT data, called CMB-SPA, gives $n_{s} =  0.9684\pm 0.0030$, which is very close to the original Planck result.  But once the recent DESI DR2 results are taken into account, the constraint on $n_{s}$ becomes very similar to the ACT constraint: $n_{s}= 0.9728\pm 0.0027$.

However, it was noted in \cite{SPT-3G:2025bzu} that combining CMB-SPA and DESI results in $\Lambda$CDM is potentially problematic, as these data sets are in significant tension with each other. In a certain sense specified in \cite{SPT-3G:2025bzu,Ferreira:2025lrd}, CMB-SPA results are $2.8 \sigma$ away from the  DESI DR2 results.  Thus, one may wonder whether it is possible to exclude any inflationary models at $2 \sigma$ by using datasets that exhibit a $2.8 \sigma$ inconsistency. 
Therefore, according to \cite{SPT-3G:2025bzu,Ferreira:2025lrd}, one should take the P-ACT-LB constraint $n_{s} =0.9743 \pm  0.0034 $ and the CMB-SPA+DESI constraint $n_{s}= 0.9728\pm 0.0027$ with caution.

\begin{figure}
\centering
\includegraphics[scale=0.23]{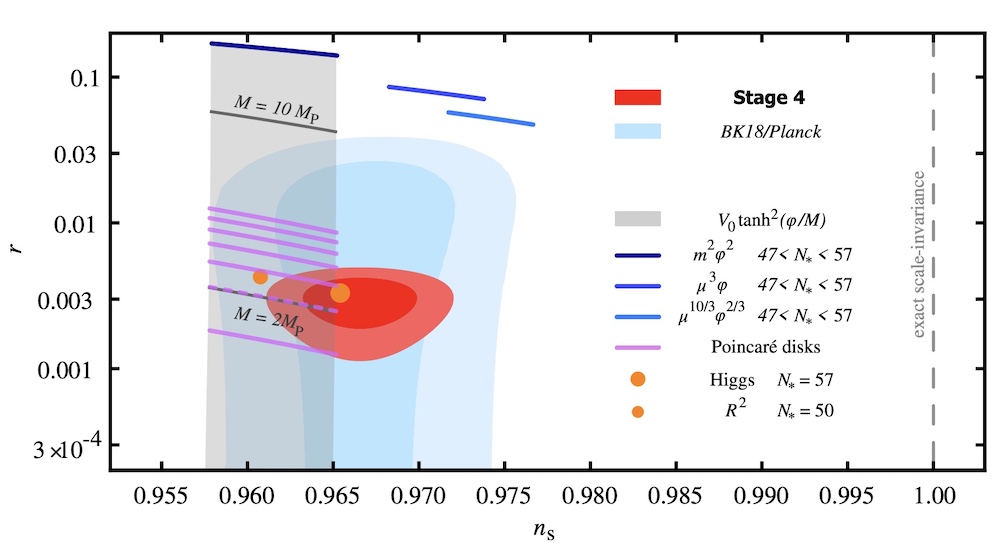}
\caption{\footnotesize Figure 2 from \cite{Chang:2022tzj} (credit R. Flaugher) for the CMB-S4 collaboration shows the predictions of  $\alpha$-attractors $\tanh^2 {\vp/\sqrt
{6\alpha}}$ (gray band), the predictions of $\alpha$-attractors for Poincar\'e disks with discrete $\alpha$ (purple lines), as well as Higgs inflation and the Starobinsky ($R^2$) model (orange circles dots)  for $47 < N_e< 57$.  Starobinsky and Higgs inflation correspond to $\alpha$-attractors with $\a=1$. Note a complete absence of targets at $n_s \gtrsim 0.966$.}
\label{Flauger0}
\end{figure}

But one should not ignore these results either, due to their potential significance. To fully appreciate it, it is sufficient to examine the targets for the B-mode searches, as shown in papers by the CMB-S4 \cite{Chang:2022tzj} and LiteBIRD \cite{LiteBIRD:2022cnt} collaborations. Here we reproduce Fig. 2 from  \cite{Chang:2022tzj}. As one can see, they show only four viable targets:  Starobinsky model, Higgs inflation, T-models of $\alpha$-attractors for arbitrary $\alpha$, and the Poincar\'e disk $\alpha$-attractors with 7 discrete values of $\alpha$.  Most importantly, these figures do not show any targets with $n_{s}$ compatible with the P-ACT-LB and CMB-SPA+DESI  constraints.

We believe it is important to be proactive and search for simple and compelling inflationary models with higher values of $n_s$.   In this paper, we will show that the simplest chaotic inflation model $\tfrac12 {m^{2}\phi^{2}}$ \cite{Linde:1983gd}  with nonminimal coupling to gravity $(1+ \phi)R$ provides a very good match to the P-ACT-LB  and the CMB-SPA+DESI constraints.

\begin{figure}
\centerline{\includegraphics[scale=.32]{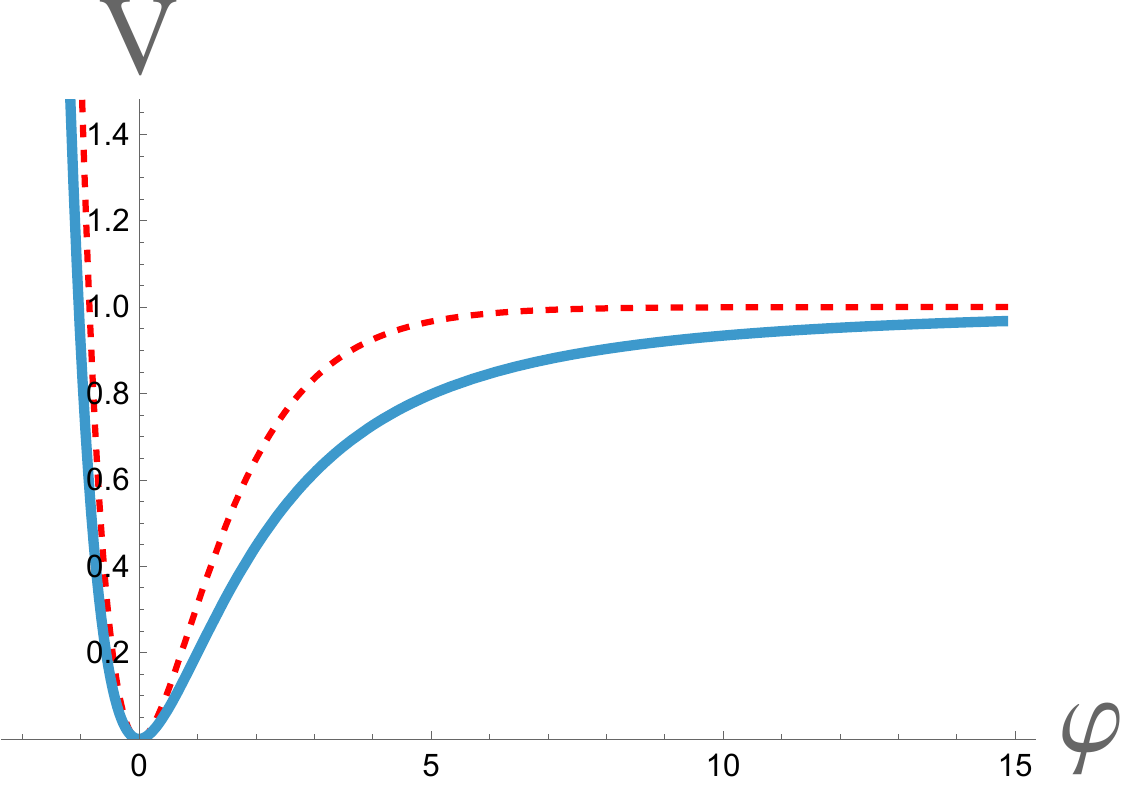}}
\caption{\footnotesize {The potential of the model \rf{Jordan} in the Einstein frame as a function of the canonically normalized inflaton field $\vp$. Unlike the potential in the Higgs inflation, this potential is not symmetric with respect to the change $\vp \to -\vp$.  Rather, it is similar to the potential in the Starobinsky model  (red dashed line) and E-models of $\alpha$-attractors \cite{Kallosh:2013yoa}, but it approaches the plateau more slowly. }}
\vspace{-.3cm}
\label{pot}
\end{figure}

The Lagrangian of this model is 
  \begin{equation}
{1\over \sqrt{-g}} \mathcal{L} =  \tfrac12 (1+\phi) R  - \tfrac12 (\partial \phi)^2 -  \tfrac12  m^{2}\phi^{2}  \ . \label{Jordan} 
 \end{equation}
One could worry that this theory is not well-defined at $\phi < -1$. However, this is not the case. 
For a simple intuitive interpretation of the theory (1), it is convenient to go to the Einstein frame by redefining the metric, $  g_{\mu \nu} \rightarrow (1+\phi)^{-1} g_{\mu \nu}$, and then switch from the field $\phi$ to the canonically normalized inflaton field $\vp$ \cite{Kallosh:2013tua}.  In the new variables, the point $\phi = -1$ corresponds to $\vp = -\infty$; the potential in the limit $\vp \to -\infty$ becomes infinitely large, as the potential in the Starobinsky model and in the E-models of $\alpha$-attractors \cite{Kallosh:2013yoa}, see Fig. \ref{pot}. Thus, the region $\phi < -1$ cannot be reached.

As one can see from this figure, the potential of the model \rf{Jordan} approaches the plateau more slowly than the potential of the Starobinsky model and the $\alpha$-attractors, which approach the plateau exponentially fast. Indeed, one can show that the potential \rf{Jordan} approaches the plateau polynomially: it is given by  
\be
V = {m^{2}\over 2} \left(1 -{8\vp^{-2}}+O(\phi^{-4})\right),
\ee
in the large field limit.

This model belongs to a general class of inflationary models previously developed in our paper  \cite{Kallosh:2013tua}:
  \begin{equation}
{1\over \sqrt{-g}} \mathcal{L}=  \tfrac12 (1+\xi f(\phi)) R  - \tfrac12 (\partial \phi)^2 -  \lambda^{2} f^{2}(\phi)  \ . \label{JordanGen} 
 \end{equation}
This theory significantly generalized the Higgs inflation model with $f(\phi) = \phi^{2}$ \cite{Salopek:1988qh,Bezrukov:2007ep}.

A detailed investigation of inflationary predictions of this class of models was performed in  \cite{Kallosh:2013tua} for  $f(\phi) = \phi^{n}$ with various values of $n$, in a broad range of $\xi$, see Fig. 1 there.  
While the focus in \cite{Kallosh:2013tua} was on the large-$\xi$ coupling, here instead we consider $\xi=n=1$ and find that, in the large-$N_e$ limit, this model predicts 
 \begin{equation}\label{nsrxi1}
n_s \approx 1- {3\over 2 N_{e}},   \qquad  r\approx {4\over \, N_{e}^{3/2}}\ .
\end{equation}
This explains an increase of $n_{s}$ and $r$ as compared to the predictions of the Starobinsky model and Higgs inflation  $n_s= 1- {2\over   N_{e}}$,  $ r={12\over  N_{e}^{2}}$, and with those of $\alpha$-attractors $n_s= 1- {2\over   N_{e}}$,  $ r={12\alpha \over  N_{e}^{2}}$  \cite{Kallosh:2013yoa}.

These analytical estimates \rf{nsrxi1} agree with the results of the numerical investigation:
 \ba\label{60}
&&n_s= 0.9733, \quad r= 0.0094, \  \qquad  N_e=60 \ ,\\
&&n_s= 0.9679, \quad r= 0.0125,  \qquad \ N_e=50 \ .
\ea
 The  $N_e=60$ predictions of this model are shown in Fig. \ref{xiStarAlone} by a yellow star close to the center of the dark purple area favored by ACT  and SPT (when combined with DESI) \footnote{Mathematica notebooks containing the derivations of these results are embedded as ancillary files in the arXiv version of this paper (2503.21030).}.
Note that tensor modes at this level ($r \sim 10^{-2}$) could be explored by BICEP/Keck \cite{BICEPKeck:2024stm} and by the Simons Observatory \cite{Hertig:2024adq}.

One can generalize our model \rf{Jordan} by introducing the parameter $\xi$ as in \rf{JordanGen},
 \begin{equation}
{1\over \sqrt{-g}} \mathcal{L}=  \tfrac12 (1+\xi \phi) R  - \tfrac12 (\partial \phi)^2 -  \tfrac12 {m^{2}}\phi^{2}  \ . \label{xi} 
 \end{equation}
In this model, one finds the following approximate expression for $n_{s}$ and $r$ in the large $N_{e}$ approximation for $\xi = O(1)$  \footnote{The same expressions for $n_s, r$ as in eq. \rf{nsrxi} appear in  pole inflation with  $p=3, c=2/\xi$ and $a_3=1/\xi $ \cite{Galante:2014ifa} and in polynomial attractors  for  $k=2 ,  \mu^2= 8/\xi$  \cite{Kallosh:2022feu}.}:
  \begin{equation}\label{nsrxi}
n_s \approx 1- {3\over 2 N_{e}},   \qquad  r\approx {4\over \xi \, N_{e}^{3/2}}\ .
\end{equation}
 Importantly, these expressions are only valid for $\xi = O(1)$,   they are not valid for $\xi \ll 1$ and for  $\xi \gg 1$. 

 The full range of values of $n_{s}$ and $r$ in this model,  for all $\xi$ from $0$ to $ \infty$, can be found in \cite{Kallosh:2013tua}, where it is given by the third trajectory from the right in Fig. 1. At  $\xi \ll 1$, the value of $r$ is too large, whereas at $\xi \gg1$ the predictions of this model coincide with the predictions of the Starobinsky model, which is disfavored by the P-ACT-LB constraints. Employing a similar numerical approach,   we found that the inflationary predictions of the model \rf{xi} for  $N_{e} = 60$ are 1$\sigma$ compatible with the P-ACT-LB constraints for order one values of the non-minimal coupling:
 \ba 
 0.3\lesssim  \xi \lesssim 4 \,.
 \ea

There are many other models with predictions compatible with the results of the latest ACT data release, such as  $D_{5}$ brane inflation \cite{Kallosh:2018zsi,Kallosh:2019hzo},  some versions of pole inflation \cite{Galante:2014ifa, Kallosh:2019hzo}, polynomial $\alpha$-attractors  \cite{Kallosh:2022feu} and hybrid $\alpha$-attractors \cite{Kallosh:2022ggf}, see a recent review \cite{Kallosh:2025ijd}.
  
The main advantage of our model \rf{Jordan}  is its ultimate simplicity.  Just like the Starobinsky model and the model $\tfrac12  m^{2}\phi^{2}$, the model \rf{Jordan} has only one parameter, $m$, which determines the amplitude of the inflationary perturbations. 

The chaotic inflation model  $\tfrac12  m^{2}\phi^{2}$ is one of the first and simplest inflationary models \cite{Linde:1983gd}. It is discussed in every textbook on inflation, but WMAP and Planck ruled it out because it predicted too large $r$. Therefore, it is quite interesting that adding a single term $\tfrac12 \phi   R$ makes the chaotic inflation model $\tfrac12  m^{2}\phi^{2}$  compatible with the latest P-ACT-LB  and  CMB-SPA+DESI data.

Moreover, its slight generalization \rf{xi} can also provide a good fit to the ACT and SPT data while allowing control of the tensor-to-scalar ratio $r$ by changing $\xi$. In this respect, the model \rf{xi} is similar to $\alpha$-attractors, where one can control $r$ by changing the parameter $\alpha$.

\smallskip 
\noindent {\bf Acknowledgements}
We acknowledge stimulating discussions with R. Bond, J.J. Carrasco, K. Ding, G. Efstathiou, F. Finelli, C. L. Kuo,  D. Meerburg,  and E. Silverstein. RK and AL are supported by the SITP and by NSF Grant PHY-2310429.

\bibliography{lindekalloshrefs}

\providecommand{\href}[2]{#2}\begingroup\raggedright\begin{thebibliography}{10}

\bibitem{Louis:2025tst}
T.~Louis {\em et al.}, ``{The Atacama Cosmology Telescope: DR6 Power Spectra,
  Likelihoods and $\Lambda$CDM Parameters}'',
  \href{http://arxiv.org/abs/2503.14452}{{\tt arXiv:2503.14452 [astro-ph.CO]}}.

\bibitem{ACT:2025tim}
{\bf ACT} Collaboration, E.~Calabrese {\em et al.}, ``{The Atacama Cosmology
  Telescope: DR6 Constraints on Extended Cosmological Models}'',
  \href{http://arxiv.org/abs/2503.14454}{{\tt arXiv:2503.14454 [astro-ph.CO]}}.

\bibitem{DESI:2024uvr}
{\bf DESI} Collaboration, A.~G. Adame {\em et al.}, ``{DESI 2024 III: Baryon
  Acoustic Oscillations from Galaxies and Quasars}'',
  \href{http://arxiv.org/abs/2404.03000}{{\tt arXiv:2404.03000 [astro-ph.CO]}}.

\bibitem{DESI:2024mwx}
{\bf DESI} Collaboration, A.~G. Adame {\em et al.}, ``{DESI 2024 VI:
  cosmological constraints from the measurements of baryon acoustic
  oscillations}'', \href{http://dx.doi.org/10.1088/1475-7516/2025/02/021}{{\em
  JCAP} {\bf 02} (2025)  021}, \href{http://arxiv.org/abs/2404.03002}{{\tt
  arXiv:2404.03002 [astro-ph.CO]}}.

\bibitem{Planck:2018vyg}
{\bf Planck} Collaboration, N.~Aghanim {\em et al.}, ``{Planck 2018 results.
  VI. Cosmological parameters}'',
  \href{http://dx.doi.org/10.1051/0004-6361/201833910}{{\em Astron. Astrophys.}
  {\bf 641} (2020)  A6}, \href{http://arxiv.org/abs/1807.06209}{{\tt
  arXiv:1807.06209 [astro-ph.CO]}}. [Erratum: Astron.Astrophys. 652, C4
  (2021)].

\bibitem{Efstathiou:2019mdh}
G.~Efstathiou and S.~Gratton, ``{A Detailed Description of the CamSpec
  Likelihood Pipeline and a Reanalysis of the Planck High Frequency Maps}'',
  \href{http://arxiv.org/abs/1910.00483}{{\tt arXiv:1910.00483 [astro-ph.CO]}}.

\bibitem{Starobinsky:1980te}
A.~A. Starobinsky, ``{A New Type of Isotropic Cosmological Models Without
  Singularity}'',
\href{http://dx.doi.org/10.1016/0370-2693(80)90670-X}{{\em Phys. Lett.} {\bf
  91B} (1980)  99--102}.

\bibitem{Salopek:1988qh}
D.~S. Salopek, J.~R. Bond, and J.~M. Bardeen, ``{Designing Density Fluctuation
  Spectra in Inflation}'',
\href{http://dx.doi.org/10.1103/PhysRevD.40.1753}{{\em Phys. Rev.} {\bf D40}
  (1989)  1753}.

\bibitem{Bezrukov:2007ep}
F.~L. Bezrukov and M.~Shaposhnikov, ``{The Standard Model Higgs boson as the
  inflaton}'', \href{http://dx.doi.org/10.1016/j.physletb.2007.11.072}{{\em
  Phys. Lett.} {\bf B659} (2008)  703--706},
\href{http://arxiv.org/abs/0710.3755}{{\tt arXiv:0710.3755 [hep-th]}}.

\bibitem{Kallosh:2013yoa}
R.~Kallosh, A.~Linde, and D.~Roest, ``{Superconformal Inflationary
  $\alpha$-Attractors}'', \href{http://dx.doi.org/10.1007/JHEP11(2013)198}{{\em
  JHEP} {\bf 11} (2013)  198},
\href{http://arxiv.org/abs/1311.0472}{{\tt arXiv:1311.0472 [hep-th]}}.

\bibitem{SPT-3G:2025bzu}
{\bf SPT-3G} Collaboration, E.~Camphuis {\em et al.}, ``{SPT-3G D1: CMB
  temperature and polarization power spectra and cosmology from 2019 and 2020
  observations of the SPT-3G Main field}'',
  \href{http://arxiv.org/abs/2506.20707}{{\tt arXiv:2506.20707 [astro-ph.CO]}}.

\bibitem{Ferreira:2025lrd}
E.~G.~M. Ferreira, E.~McDonough, L.~Balkenhol, R.~Kallosh, L.~Knox, and
  A.~Linde, ``{The BAO-CMB Tension and Implications for Inflation}'',
  \href{http://arxiv.org/abs/2507.12459}{{\tt arXiv:2507.12459 [astro-ph.CO]}}.

\bibitem{Chang:2022tzj}
C.~L. Chang {\em et al.}, ``{Snowmass2021 Cosmic Frontier: Cosmic Microwave
  Background Measurements White Paper}'',
  \href{http://arxiv.org/abs/2203.07638}{{\tt arXiv:2203.07638 [astro-ph.CO]}}.

\bibitem{LiteBIRD:2022cnt}
{\bf LiteBIRD} Collaboration, E.~Allys {\em et al.}, ``{Probing Cosmic
  Inflation with the LiteBIRD Cosmic Microwave Background Polarization
  Survey}'', \href{http://dx.doi.org/10.1093/ptep/ptac150}{{\em PTEP} {\bf
  2023} (2023) no.~4, 042F01}, \href{http://arxiv.org/abs/2202.02773}{{\tt
  arXiv:2202.02773 [astro-ph.IM]}}.

\bibitem{Linde:1983gd}
A.~D. Linde, ``{Chaotic Inflation}'',
\href{http://dx.doi.org/10.1016/0370-2693(83)90837-7}{{\em Phys. Lett.} {\bf
  B129} (1983)  177--181}.

\bibitem{Kallosh:2013tua}
R.~Kallosh, A.~Linde, and D.~Roest, ``{Universal Attractor for Inflation at
  Strong Coupling}'',
  \href{http://dx.doi.org/10.1103/PhysRevLett.112.011303}{{\em Phys. Rev.
  Lett.} {\bf 112} (2014) no.~1, 011303},
\href{http://arxiv.org/abs/1310.3950}{{\tt arXiv:1310.3950 [hep-th]}}.

\bibitem{Note1}
Mathematica notebooks containing the derivations of these results are embedded
  as ancillary files in the arXiv version of this paper (2503.21030).

\bibitem{BICEPKeck:2024stm}
{\bf BICEP/Keck} Collaboration, P.~A.~R. Ade {\em et al.}, ``{Constraining
  Inflation with the BICEP/Keck CMB Polarization Experiments}'', in {\em {58th
  Rencontres de Moriond on Cosmology}}.
\newblock 5, 2024.
\newblock \href{http://arxiv.org/abs/2405.19469}{{\tt arXiv:2405.19469
  [astro-ph.CO]}}.

\bibitem{Hertig:2024adq}
E.~Hertig, K.~Wolz, T.~Namikawa, A.~Baleato~Lizancos, S.~Azzoni, and
  A.~Challinor, ``{The Simons Observatory: Combining delensing and foreground
  cleaning for improved constraints on inflation}'', in {\em {58th Rencontres
  de Moriond on Cosmology}}.
\newblock 5, 2024.
\newblock \href{http://arxiv.org/abs/2405.13201}{{\tt arXiv:2405.13201
  [astro-ph.CO]}}.

\bibitem{Note2}
The same expressions for $n_s, r$ as in eq. (\ref {nsrxi}) appear in pole
  inflation with $p=3, c=2/\xi $ and $a_3=1/\xi $ \cite {Galante:2014ifa} and
  in polynomial attractors for $k=2 , \mu ^2= 8/\xi $ \cite {Kallosh:2022feu}.

\bibitem{Kallosh:2018zsi}
R.~Kallosh, A.~Linde, and Y.~Yamada, ``{Planck 2018 and Brane Inflation
  Revisited}'', \href{http://dx.doi.org/10.1007/JHEP01(2019)008}{{\em JHEP}
  {\bf 01} (2019)  008}, \href{http://arxiv.org/abs/1811.01023}{{\tt
  arXiv:1811.01023 [hep-th]}}.

\bibitem{Kallosh:2019hzo}
R.~Kallosh and A.~Linde, ``{CMB targets after the latest $Planck$ data
  release}'', \href{http://dx.doi.org/10.1103/PhysRevD.100.123523}{{\em Phys.
  Rev.} {\bf D100} (2019) no.~12, 123523},
\href{http://arxiv.org/abs/1909.04687}{{\tt arXiv:1909.04687 [hep-th]}}.

\bibitem{Galante:2014ifa}
M.~Galante, R.~Kallosh, A.~Linde, and D.~Roest, ``{Unity of Cosmological
  Inflation Attractors}'',
  \href{http://dx.doi.org/10.1103/PhysRevLett.114.141302}{{\em Phys. Rev.
  Lett.} {\bf 114} (2015) no.~14, 141302},
\href{http://arxiv.org/abs/1412.3797}{{\tt arXiv:1412.3797 [hep-th]}}.

\bibitem{Kallosh:2022feu}
R.~Kallosh and A.~Linde, ``{Polynomial \ensuremath{\alpha}-attractors}'',
  \href{http://dx.doi.org/10.1088/1475-7516/2022/04/017}{{\em JCAP} {\bf 04}
  (2022) no.~04, 017}, \href{http://arxiv.org/abs/2202.06492}{{\tt
  arXiv:2202.06492 [astro-ph.CO]}}.

\bibitem{Kallosh:2022ggf}
R.~Kallosh and A.~Linde, ``{Hybrid cosmological attractors}'',
  \href{http://dx.doi.org/10.1103/PhysRevD.106.023522}{{\em Phys. Rev. D} {\bf
  106} (2022) no.~2, 023522}, \href{http://arxiv.org/abs/2204.02425}{{\tt
  arXiv:2204.02425 [hep-th]}}.

\bibitem{Kallosh:2025ijd}
R.~Kallosh and A.~Linde, ``{On the Present Status of Inflationary Cosmology}'',
  \href{http://arxiv.org/abs/2505.13646}{{\tt arXiv:2505.13646 [hep-th]}}.

\end{thebibliography}\endgroup

\bibliographystyle{utphys}

\end{document}